\documentclass[12pt,preprint]{aastex}
\begin{document}

\parindent=1.0cm

\title{The Lop-Sided Spiral Galaxy NGC 247: Clues to a Possible 
Interaction with NGC 253}

\author{T.J. Davidge\altaffilmark{1,2}}

\affil{Dominion Astrophysical Observatory,
\\Herzberg Astronomy \& Astrophysics Research Center,
\\National Research Council of Canada, 5071 West Saanich Road,
\\Victoria, BC Canada V9E 2E7\\tim.davidge@nrc.ca; tdavidge1450@gmail.com}

\altaffiltext{1}{This research has made use of the NASA/IPAC Infrared Science 
Archive, which is funded by the National Aeronautics and Space Administration and 
operated by the California Institute of Technology.}

\altaffiltext{2}{This research has made use of the NASA/IPAC Extragalactic Database 
(NED), which is operated by the Jet Propulsion Laboratory, California Institute of 
Technology, under contract with the National Aeronautics and Space Administration.}

\begin{abstract}

	Observations that span a broad range of wavelengths are used to examine 
asymmetries in the disk of the nearby late-type spiral galaxy NGC 247. 
The northern spiral arm is over-luminous at all wavelengths when compared 
with other parts of the galaxy at similar galactocentric radii, while the density 
of very luminous red stars in the void that is immediately south of this arm matches 
that in other parts of the disk at the same galactocentric radius. 
Two bubbles with spatial extents of many kpc are identified in the disk, and 
many of the young stars in the southern disk of NGC 247 are located in the 
walls of one of these structures. Dynamical age estimates of these bubbles 
coincide with the last large-scale star formation event in the nucleus, 
suggesting that there was large-scale star formation throughout the disk of NGC 247 
a few hundred Myr in the past. Morphological similarities 
are seen with the classical lop-sided galaxy NGC 4027, and it is concluded that 
NGC 247 is a significantly lop-sided spiral galaxy. 
The void in the northern disk is then the area between 
the main body of the disk and the northern arm viewed in projection. The implications
of a lop-sided morphology for NGC 247 in the context of interactions with its nearby 
starburst galaxy companion NGC 253 are discussed.

\end{abstract}

\section{INTRODUCTION}

	Nearby galaxies are unique laboratories for examining the processes that shape 
galaxy evolution. NGC 247 is a nearby SAB(s)d (de Vauncouleurs et al. 1991) galaxy. 
While it is one of many seemingly non-descript late-type disk galaxies located within 
a few Mpc of the Galaxy, there are hints that NGC 247 has had an eventful past, and 
so might be of special interest. For example, there are hints that NGC 247 has 
interacted with another galaxy. The HI disk of NGC 247 is compact (Carignan \& 
Puche 1990), and this characteristic is seen among galaxies in 
crowded environments (e.g. Chung et al. 2009). 
The stellar disk of NGC 247 also extends well beyond the HI disk (Davidge 
2006), and spatially extended stellar disks may be the result of angular momentum 
transfer via interactions with other galaxies (e.g. Hammer et al. 2007). 
While galaxy-galaxy interactions are not in themselves rare events, NGC 247 is 
a companion of the much more extensively studied late-type spiral galaxy NGC 253 
(Karachentsev et al. 2003), which is one of the nearest large starburst galaxies.
The relationship between NGC 247 and NGC 253 might then provide clues into the 
cause of the starburst in the latter galaxy.

	In addition to the evidence of a possible interaction, the recent specific star 
formation rate (sSFR) of NGC 247 suggests that it has been quiescent 
when compared with other spiral galaxies of similar mass. Based on the 
properties of bright main sequence stars, Davidge (2006) found that the most 
recent episode of star formation ended $\sim 6$ Myr in the past, and that 
the recent SFR is $\sim 0.1$ M$_\odot$ year$^{-1}$. This SFR places NGC 247 
almost an order of magnitude below the main sequence of galaxies on the 
sSFR $vs$ stellar mass plane presented by Popesso et al. (2019). 

	The central regions of galaxies are the deepest part 
of the gravitational potential well and so -- in the absence of feedback -- 
are areas where star formation might continue after 
activity in the disk has been quenched. Galaxy nuclei 
may then harbor supplemental information about the past history of the host.
Davidge \& Courteau (2002) find that the $J-K$ color of the NGC 247 nucleus is 
not remarkable. While the $J-K$ color is redder than that of the circumnuclear 
surroundings and the nucleus of NGC 2403, it is similar to that of the center of M33. 
Kacharov et al. (2018) conclude from an analysis of blue and visible wavelength 
spectra that the nucleus of NGC 247 is predominantly old, with the bulk of stellar 
mass forming $> 1$ Gyr ago. Still, the nucleus has not been quiescent during 
the past Gyr, as there is evidence for an uptick in 
the nuclear SFR $0.1 - 0.5$ Gyr in the past. Curiously, Kacharov et al. (2018) also 
find that the stars that formed at that time have super-solar metallicities. 

	The possible presence of stars with super-solar 
metallicities in the nucleus of NGC 247 is intriguing, as 
the integrated $K$ brightness of NGC 247 (M$_K \sim -20.5$ assuming a distance 
modulus of 27.9 -- see below, with $K_{tot} = 7.4$; Jarrett et al. 2003) is similar 
to that of M33. Young stars in NGC 247 should then have an M33-like (i.e. sub-solar; 
Rosolowsky \& Simon 2008) metallicity if NGC 247 falls along the metallicity $vs.$
stellar mass relation defined by other galaxies (e.g. Saviane et al. 2008). Barring 
a peculiar chemical enrichment scenario for the center of NGC 247, then a  
super-solar metallicity for stars in this part of the galaxy might 
require a special origin for the gas from which they formed. 

	The morphological properties of a galaxy also 
yield insights into its past. The nucleus of NGC 247 is offset from the 
geometric center as defined by isophotes near the edge of the disk. NGC 247 also has a 
conspicuous asymmetry in the form of an area of low stellar density (hereafter referred 
to as the `void') in the northern part of the disk (e.g. Carignan 1985). 
Finally, the rotation curve of NGC 247 is asymmetric along the major 
axis (Carignan \& Puche 1990). 

	The nature of the void and other asymmetries in the disk are of obvious
importance for understanding the evolution of the galaxy. Wagner-Kaiser et al. 
(2014) examine the stellar content in the void and find an absence of 
bright blue stars, suggesting a lull in recent star formation 
during at least the past Gyr. They find that the void contains luminous red stars, 
signalling that a substrate of older stars is present. Referencing HI maps 
presented by Ott et al. (2012) and Warren et al. (2012), 
Wagner-Kaiser et al. (2014) note that the void also has a lower HI 
density than its surroundings, although there are areas with similar HI densities at 
other locations in the disk. Wagner-Kaiser et al. (2014) discuss mechanisms 
that might produce the void, and propose that it is a region of low gas content
that is the result of the recent passage of a $\sim 10^8$M$_\odot$ 
dark subhalo that contains gas. 

	In the present study, the structure of the NGC 247 disk is examined using 
archival images that cover ultraviolet (UV) and mid-infrared (MIR) wavelengths. 
These images trace light from stars that formed during a broad range of epochs, 
including when NGC 247 may have experienced events that shaped its evolution. 
Efforts to study the morphology of NGC 247 are complicated 
by its oblique orientation on the sky. Therefore, the images 
are deprojected to provide a panoramic view of the 
galaxy, thereby facilitating the exploration of asymmetries in the disk.

	During the past two decades a number of studies 
have examined the distance of NGC 247, producing very similar 
results. Observations of Cepheids over a broad range of wavelengths 
yield a distance modulus between 27.6 and 28.0 (Garcia-Varela et al. 2008; Madore 
et al. 2009; Gieren et al. 2009), with the majority of measurements favoring 
a distance modulus of 27.8 (Gieren et al. 2009). For comparison, stars near the giant 
branch tip yield a distance modulus of $\sim 27.9$ (Davidge 2006; Karachentsev et al. 
2006). The distance based on the giant branch tip is adopted for the present work, 
placing NGC 247 at a distance of 3.8 Mpc.

	The paper is structured as follows. Details of the observations that serve 
as the basis for this study are presented in Section 2. Deprojected images of 
NGC 247 are used to examine the large-scale photometric 
characteristics of asymmetric structures in Section 3. 
The distribution of partially resolved objects in the SPITZER IRAC [3.6] 
and [4.5] images are discussed in Section 4, where the projected 
density of objects in the northern void is compared with that in other parts 
of the galaxy at similar galactocentric radii. 
Comparisons are made with a classic lop-sided spiral galaxy in Section 5, 
and it is concluded that NGC 247 is an overt lop-sided spiral galaxy. 
A discussion and summary of the results follows in Section 6.

\section{ARCHIVAL IMAGES}

	Processed archival images recorded with the GALEX (Martin et al. 2005), WISE 
(Wright et al. 2010), and SPITZER (Werner et al. 2004)
satellites are used in the present study. The SPITZER and WISE images were 
downloaded from the IRSA website \footnote[3]
{https://irsa.ipac.caltech.edu/Missions/}. 
The GALEX observations were downloaded from the Nasa Extragalactic 
Database (NED) \footnote[4]{https://ned.ipac.caltech.edu/classic/}.

	The GALEX images are from the Atlas of Nearby Galaxies 
(Gil de Paz et al. 2007). The FUV images are most sensitive to massive stars 
that formed within the past few Myrs, while the NUV 
images are more sensitive to lower mass stars that formed 
a few hundred Myr ago. When considered together, the FUV and NUV images 
thus probe light from stars that formed from the present day to the epoch that Kacharov 
et al. (2018) identify with recent star formation in the nucleus of NGC 247.

	The WISE images are from the All-sky survey (Wright et al. 2010). 
The light that dominates the W1 and W2 images originates 
predominantly in luminous red stars that formed 
during early or intermediate epochs; hence, the W1 and W2 images 
trace stellar mass. In contrast, much of the light sampled in the W3 
and W4 images may originate from thermal emission, and in late-type spiral galaxies 
such as NGC 247 much of the emission at these wavelengths comes from dust 
that is heated by massive main sequence stars (e.g. Cluver et al. 2014) 
and/or post-asymptotic giant branch (AGB) stars. While much the same can be said 
for W4 observations, the W3 images detect fainter structures 
than those in W4 and have better angular resolution. Given that the depth and image 
quality of the W1 and W3 images are superior to those in W2 and W4, it was 
decided to consider only the W1 and W3 images for this study.

	The SPITZER IRAC (Fazio et al. 2004) recorded
[3.6] and [4.5] images of NGC 247 as part of the SPIRITS Survey (Kasliwal et al. 2017). 
The galaxy was observed a number of times, and the images cover a 
range of epochs and on-sky orientations. However, 
only two pointings include the void and the northern spiral arm that are important 
targets for the present study. These have 
astronomical observation request (AOR) identifications 
50548736 and 50548992, and only images from these AORs are considered here. 

\section{INTEGRATED LIGHT IN THE UV AND IR}

	NGC 247 is inclined to the line-of-sight, and images were generated to 
simulate the face-on appearance of the galaxy. This was done by assuming that the 
disk is infinitely thin and restricted to a single plane. 
An inclination of 74$^o$ was adopted, as deduced from HI observations 
by Carignan \& Puche (1990). 

	With the assumptions described above, a deprojected image can be 
constructed by stretching the disk along the minor axis by an amount defined by 
the inclination. Circular isophotes will result if the disk is symmetric 
and the inclination is correct. This procedure is 
passive in nature, as signal is neither added to nor subtracted from 
the original image. Rather, the manner in which 
information in the images is presented is altered by compressing the frequency 
distribution along one axis in a linear (i.e. reversible) way. The structures 
discussed here are present in the original images, although they tend to 
be less obvious in those images due to the orientation of NGC 247 
on the sky. Deconvolution filters can be applied to assess 
distortions introduced by deprojection, and this is discussed in the Appendix. 

	There are inherent limitations to 
deprojecting a three-dimensional structure using two-dimensional information. 
The deprojection procedure used here does not account for three dimensional features 
such as warps that might result from -- say -- tidal interactions. 
Warps will smear the deprojected image, with asymmetries 
in the overall structure being one possible result. Still, 
there are indications that warping is not a factor in NGC 247. Carignan \& 
Puche (1990) do not find evidence of warping in HI observations of the 
outer disk of NGC 247, although they do find that the rotation curve is 
highly asymmetric along the major (roughly north-south) axis. Hlavacek-Larrondo et al. 
(2011) also do not find evidence of warping in Fabry-Perot H$\alpha$ observations 
of NGC 247.

	The finite thickness of disks is another factor 
that could introduce smearing, and the extent of this smearing becomes 
progressively more important as one moves to larger inclination angles. However, 
stars with different ages have different characteristic distance dispersions from the 
disk plane, and so the degree of smearing depends on the age of the population 
being studied. Images that sample younger stars, such as those recorded through 
the FUV or NUV filters, should be less susceptible to disk thickness 
effects than -- say -- the W1 image. A comparison of images that sample populations 
with very different ages then have the potential to monitor smearing 
induced by disk thickness; structures that appear in deprojected images 
that span wide wavelength ranges will likely not be artifacts of disk thickness. 
An obvious complication is that the spectral energy distributions of 
stellar populations will also introduce differences between images 
that span a broad wavelength range.

	Observed and deprojected images of NGC 247 are shown 
in Figures 1 (FUV and NUV images) and 2 (W1 and W3 images). 
The 8th magnitude foreground star 2MASS00470164-2052106 is seen 
against the southern disk of NGC 247 and is a prominent feature in the W1 and W3 
images. Light from this star was subtracted from the images prior to 
deprojection. Low level noise artifacts remain, 
and the area that contain these is indicated with a blue box in Figure 2.

\begin{figure}
\figurenum{1}
\epsscale{1.0}
\plotone{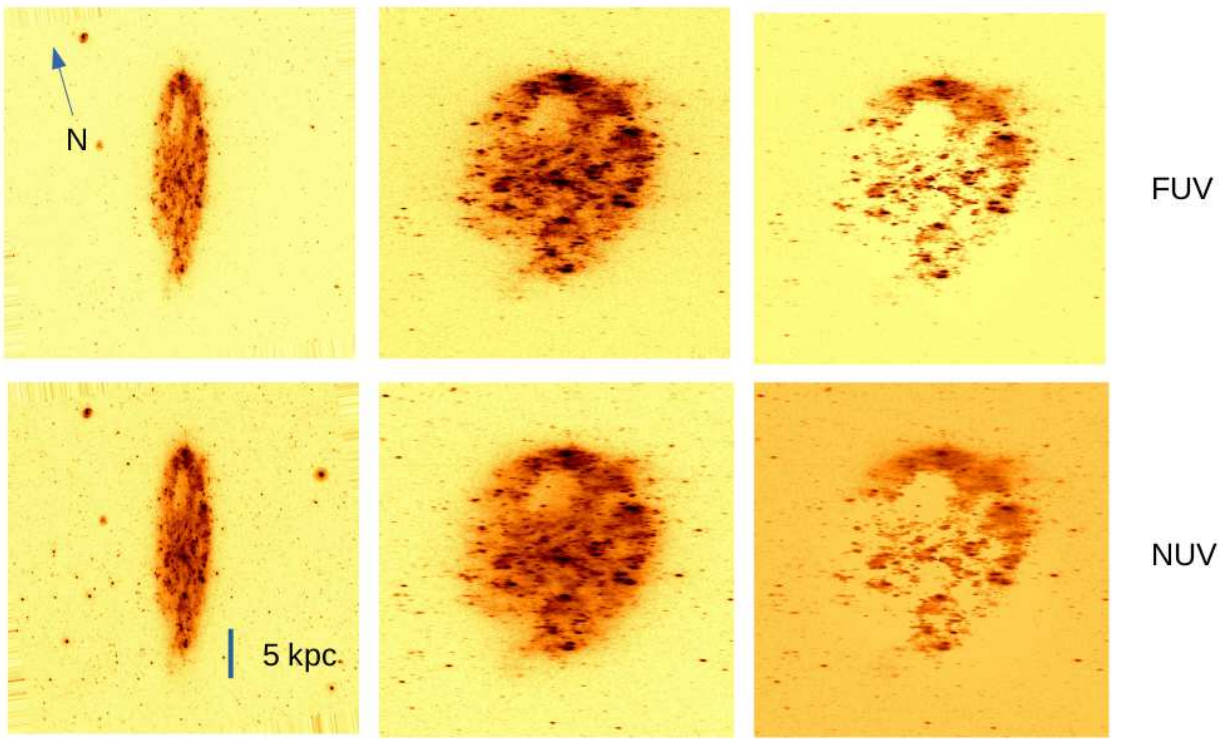}
\caption{GALEX FUV and NUV images of NGC 247, rotated so that the major axis 
is vertical. North is indicated in the upper left hand panel. As with all 
images shown throughout this paper, the intensity map has been inverted to produce 
a negative image. The left hand column shows the observed images, 
while the middle column shows the images deprojected 
with an assumed inclination of 74$^o$. The right hand column shows the deprojected 
images after the subtraction of an azimuthally symmetric model of diffuse light. 
The northern void is seen in both filters. The northern spiral arm is 
the dominant structure after the removal of the smooth disk component, highlighting 
the lop-sided nature of the galaxy. In addition to the northern void, there is also 
a region with little or no FUV and NUV signal at the southern edge of the disk. 
Prominent ring-shaped structures are also seen in the deprojected images.}
\end{figure}

\begin{figure}
\figurenum{2}
\epsscale{1.0}
\plotone{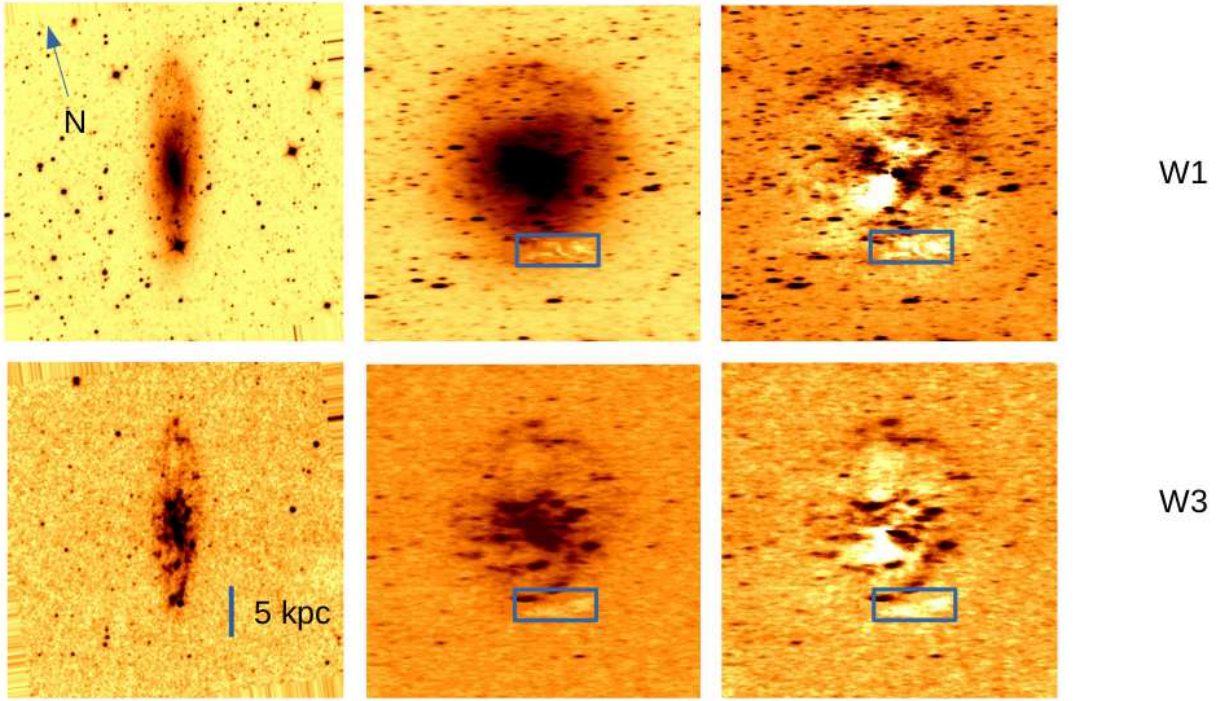}
\caption{Same as Figure 1, but showing images in the W1 and W3 filters.
Light from the 8th magnitude star 2MASS00470164-2052106 was subtracted 
from these images prior to deprojection, and the 
area that contains residuals from this star is indicated with 
the blue box in each panel. This box is elongated because 
deprojection distorts the PSF along one axis. The distribution 
of light in the W1 images is smoother than in W3, 
as expected given the spatial distribution of the old and intermediate-age evolved 
stars that are a major contributor to the light in W1. The northern void is a 
subtle feature in the deprojected W1 image, and it 
has a background-like surface brightness in the deprojected 
W3 image. The northern spiral arm stands out in the disk-subtracted images in 
both filters, highlighting the lop-sided character of NGC 247. The void near the 
southern edge of the disk that is conspicuous in the GALEX images is not as obvious 
in the IR. The sources at small radii in the W3 image may be concentrations 
of luminous AGB stars that formed at the same time as the recent episode of 
nuclear star formation that was characterised by Kacharov et al. (2018).}
\end{figure}

\begin{figure}[!ht]
\figurenum{3}
\epsscale{1.0}
\plotone{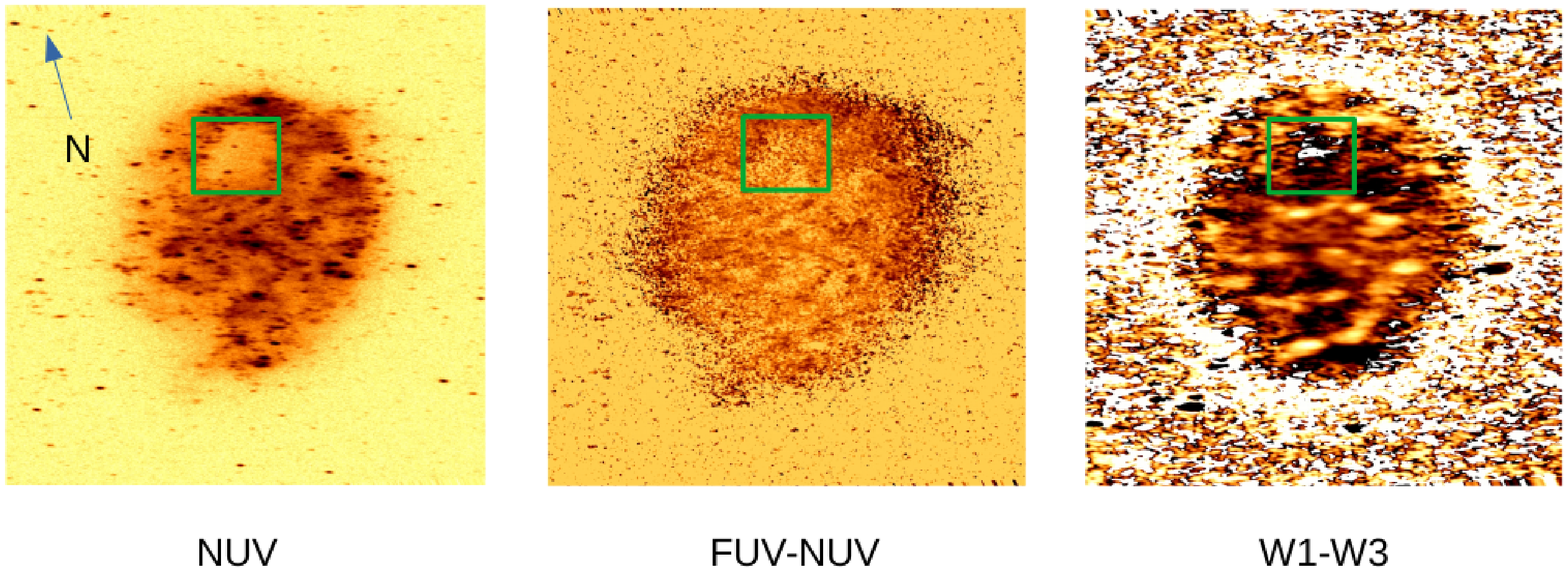}
\caption{UV and IR colors in the deprojected disk of NGC 247. Lighter shading 
indicates redder colors. The deprojected NUV image from Figure 1 is shown to aid in 
the identification of structures, and the void is marked 
with a green box in each panel. The northern void 
has a UV color that is not remarkable when compared with other areas in the 
disk. The void has a blue color in the W1--W3 map, and the white area in the 
void is likely due to low signal in the W3 image. The northern arm 
has blue UV colors and red W1--W3 colors, and this is consistent with it 
being an area that contains very hot stars that are heating nearby dust.}
\end{figure}

\subsection{Diffuse UV Light}

	Much of the light in the GALEX FUV and NUV images originates from young, 
massive stars, and these define prominent structures in the deprojected 
UV images. There is a concentration of bright sources 
near the northern and eastern edge of the spiral arm that defines 
the northern edge of the disk (hereafter the 'northern arm'). There is also a 
a serpentine-like structure that winds its way south of the 
nucleus that is also prominent in the distribution of young stars shown in Figure 8 
of Rodriguez et al. (2009). Deprojection reveals that this structure appears to be 
the walls of circular bubbles, which are discussed at greater length in the next 
section. 

	The northern void has a surface brightness in the UV images 
that is markedly lower than its surroundings and is close to background 
values. Still, the presence of diffuse light in the northern void in 
the deprojected NUV image suggests that it may contain stars that formed a few 
hundred Myr in the past. Main sequence turn-off (MSTO) stars of this age fall 
below the faint limit of the images discussed by Walter-Kaiser et al. (2014). 
The northern void is also not unique, as there is a notch-like area at the edge of 
the southern disk that is deficient in hot stars.

	A map of FUV--NUV colors generated from the deprojected GALEX images is 
shown in the middle column of Figure 3. The northern arm has 
blue FUV--NUV colors that extend over much of its angular extent, highlighting 
it as an area of very recent star formation. The northern void has a 
FUV--NUV color that is not unique in the NGC 247 disk, although UV 
colors are sensitive to variations in extinction. 

	A radially symmetric light profile was constructed by combining 
azimuthally the light in each of the deprojected GALEX images. 
The nucleus was taken to be the true center of the galaxy, and 
the median signal at each radius was found. The median, rather than 
the mean, was taken at each radius to suppress the discrete structures that dominate 
the UV images. The disk profile constructed solely 
from the southern half of the disk, where there are fewer aymmetries 
than in the northern half of the disk, is 
similar to that obtained from the entire disk. 
While a model of diffuse disk light could be extracted from a conventional isophotal 
analysis, the intent of the current work is to examine asymmetric structures in 
the disk, justifying the use of the azimuthal smoothing technique employed here.

	The results of subtracting the diffuse disk component from the deprojected 
FUV and NUV images are shown in the right hand column of Figure 1. 
The asymmetric distribution of recent star formation sites in NGC 247 is 
highlighted with the removal of diffuse light. The northern arm 
is the dominant UV structure in the disk-subtracted images, emphasizing the 
lop-sided nature of NGC 247 at these wavelengths.

\subsubsection{Rings in the Disk}

	Prominent large-scale bubbles are seen in the deprojected UV 
images, and two of these are examined in Figure 4. The circular shape of these 
bubbles is suggestive of expansion into a highly uniform interstellar medium 
(ISM). One of the rings is near the base of the northern arm, and is adjacent to 
the void. This ring has a diameter of $\sim 100$ arcsec, or $\sim 2$ kpc, 
placing it within the size range found in dwarf galaxies (e.g. Pokhrel et al. 2020). 
The other ring is in the southern part of the disk, and 
has a diameter of $\sim 180$ arcsec, or $\sim 3.3$ kpc, making it 
larger than bubbles found in dwarf galaxies. 

\begin{figure}
\figurenum{4}
\epsscale{1.0}
\plotone{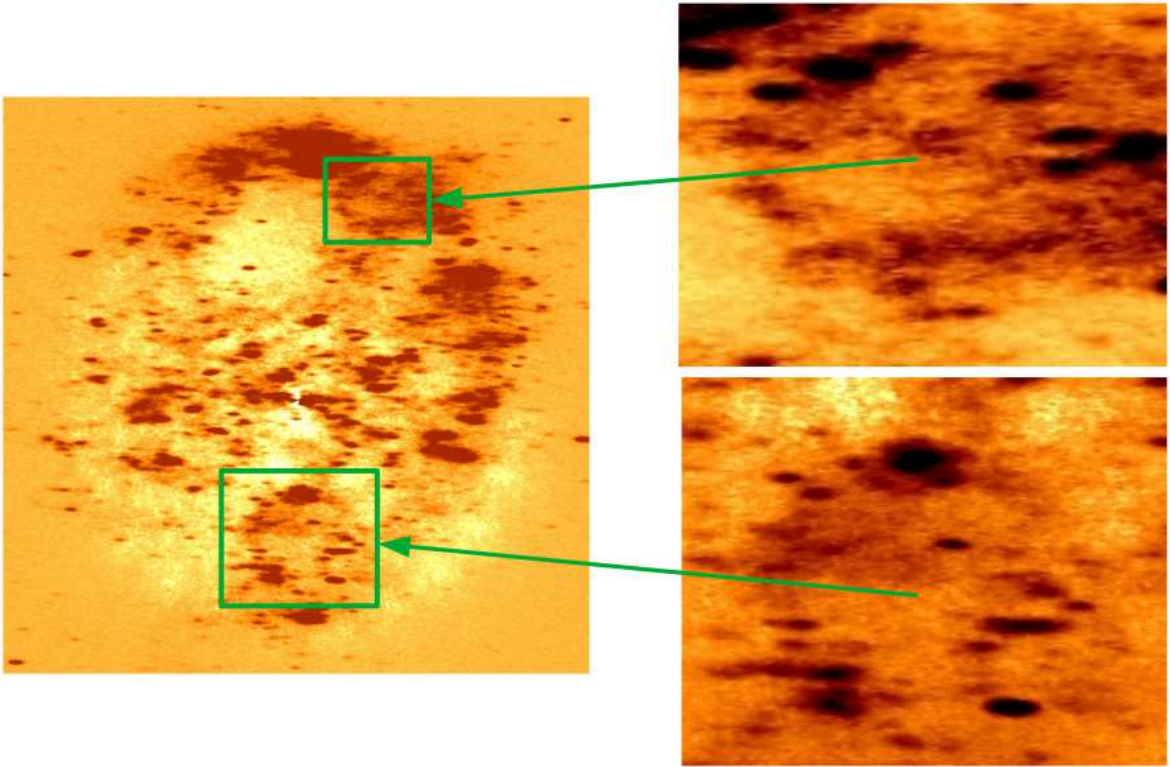}
\caption{Two large-scale bubbles in the disk of NGC 247. The deprojected 
disk-subtracted NUV image of NGC 247 is shown to the left, 
while the areas centered on the northern (top) and southern (bottom) 
bubbles are shown to the right. Sources are seen inside both 
bubbles, although none are near the geometric centers. 
There is a less well-defined bubble immediately above the southern 
bubble in the disk-subtracted NUV image.}
\end{figure}

	The light in the rings orignates from diffuse nebular emission and 
light from discrete sources, with the latter including 
bright stars, HII regions, and compact star clusters. 
Individual sources are seen inside the bubbles, although none are near the 
geometric centers of the rings. Both rings intersect with 
concentrations of bright blue stars in Figure 8 of Rogriguez et al. (2019), as 
well as with H$\alpha$ emission sources in Ferguson et al. (1995).
Neither ring stands out in the W1 or W3 images. This is not unexpected, since even 
though discrete hot sources will power hot dust emission that could 
be detected in the W3 image, on their own discrete sources may not define an obvious 
ring structure. Neither ring stands out in the HI maps discussed by Braun (1995). 

	The southern ring is part of an extensive distribution of young 
stars. Figure 8 of Rodriguez et al. (2009) shows the 
distribution of bright stars in the southern half of NGC 247, and the majority of 
these define a serpentine-like structure. The lower part of this structure 
forms one arc of the southern bubble shown in Figure 4. The bright stars to the north of 
this form a counter-arc in the deprojected image, and there is a less 
well-defined bubble immediately above the southern bubble 
in the left hand panel of Figure 4. The diameter of this other bubble 
is similar to that of its southern neighbor, suggesting a similar age.

	Dynamical age estimates for the rings can be found if the expansion velocity 
is known. A caveat is that a single expansion velocity likely does not apply 
throughout the lifetime of a bubble, since the bubble 
shock front will encounter inhomogeneities in the ISM as 
it expands. Still, Puche et al. (1992) examine superbubbles in the dwarf galaxy 
Holmberg II, and their HI observations reveal structures that span a 
range of sizes and ages. The dispersion in expansion velocities is only a 
few km/sec, suggesting that variations in expansion velocity with age may not 
be substantial in that galaxy. The largest bubbles 
in their sample have expansion velocities of $\sim 7$ km/sec. 
Applying this velocity to the southern bubble in NGC 247 yields 
an age of 230 Myr, while for the northern bubble an age of 150 Myr is found.

	These age estimates fall within the range of ages estimated for the 
latest episode of star forming activity in the nucleus of NGC 247 by 
Kacharov et al. (2018). The bubbles in Figure 4 may then have 
formed as part of a large-scale episode of star formation that 
also included the galaxy nucleus. In fact, one model for bubble formation proposes 
that they are the result of outflows from concentrated areas of star 
formation, with star formation induced in the bubble as a shock front moves 
outwards (e.g. McCray \& Kafatos 1987). This mechanism satisfies the energy 
balance requirements needed to produce bubbles in dwarf galaxies (Warren et a. 2011). 

	There are bubbles, including those in the 
NGC 247 disk and many of those studied by Warren 
et al. (2011), that lack an obvious central OB association or HII region. 
This could simply be because the central source has aged and faded with 
time. The age estimates for the NGC 247 bubbles, as well as many of those 
studied by Puche et al. (1992), are consistent with such an explanation. 
Still, it has also been suggested that bubbles 
may result from the infall of high velocity clouds (HVCs; e.g. Park et al. 2016). 
An attractive feature of interactions with HVCs in the case of NGC 247 is that 
they may deliver material to the galaxy that does not reflect its chemical enrichment 
history. An infall of material stripped from another galaxy -- notionally 
NGC 253 in the case of NGC 247 -- might then provide a means 
of delivering metal-enriched material into the regions near 
the NGC 247 nucleus, thereby explaining the presence of the metal-rich population 
found by Kacharov et al. (2018).

\subsection{Diffuse IR Light}

	The inner disk of NGC 247 in the deprojected W1 image has a 
comparatively smooth light distribution that is centered on the galaxy nucleus and 
extends over $\sim 11$ arcminutes. The central few 
arcminutes of the W1 image have a boxy morphology that includes a central bar.
The northern spiral arm is also a prominent feature in W1, reinforcing 
the visual impression of NGC 247 as a lop-sided spiral galaxy. 

	The surface brightness of the void in the deprojected W1 image is higher 
than in the background, and is not greatly different from that 
in other parts of the disk at similar galactocentric radii. This similarity 
in red stellar density is demonstrated further in Section 4 using star counts. 
The void near the southern edge of the disk that was identified in the FUV and NUV 
images is not evident in the W1 image, indicating that this feature 
is populated by red stars, like its northern counterpart. 

	While there is expected to be some similarity between the UV and W3 images, 
the diffuse component in the W3 image in Figure 2 is more centrally-concentrated than 
in the UV, as expected if the light originates from stars that span a broader age range 
than in the UV. If there was large scale star formation in 
the NGC 247 disk a few hundred Myr in the past (Section 3.1.1), then some of the sources 
in the W3 image may be dust-enshrouded AGB stars. 
Much of the light in the W3 image is in the southern part of the disk.

	W1--W3 colors in NGC 247 are shown in the right hand 
panel of Figure 3. There are sources with red W1--W3 colors in the northern spiral 
arm, and this is consistent with thermal emission from hot dust that 
is heated by young stellar groupings. The northern void also contains pockets of 
localized blue W1--W3 color, although the W3 signal is weak there, 
making integrated light measurements uncertain. In fact, the white area  
in the void in the W1--W3 color map corresponds to an area of 
low W3 signal. Still, the overall W1--W3 color of the void is not 
exceptional, as there are other areas in the disk that have a similar color. 

	The result of subtracting a radially-smoothed light profile from the W1 
image is shown in the right hand column of Figure 2. 
The residuals near the galaxy center are due to the removal of a 
symmetric light profile from a region that contains a bar. 
There are also negative residuals in the northern void 
in the disk-subtracted W1 image in Figure 2. While seemingly at odds with the void 
W1 surface brightness noted earlier in this section as well as with star counts 
(Section 4), we suspect that the low surface brightness in the disk-subtracted 
W1 image may be an artifact of the radial smoothing procedure in the presence of 
asymmetries. In fact, a similar result is seen in the 
space between a tidal arm and the main disk body of a classic lop-sided galaxy that is 
similar in appearance to NGC 247 (Section 5). The prominent nature and large 
angular extent of the northern spiral arm is consistent with NGC 247 being a 
lop-sided spiral galaxy. 

	The residuals in the disk-subtracted W3 image highlight the star-forming 
regions and luminous AGB stars in the NGC 247 disk. While the 
northern spiral arm is prominent after the removal of diffuse light, 
it is the sources in the southern part of the disk that
are the most luminous in the W3 image. The northern void in the disk-subtracted 
W3 image also has a lower than average surface brightness for that radius, in 
agreement with what is seen in the UV. 

\subsubsection{The Smoothed Infrared Light Profile}

	Carignan (1985) examined the light profile of NGC 247 at blue and 
red wavelengths. The profiles at both wavelengths show a central cusp that extends
over 2 -- 3 kpc in radius, and a disk component that can be traced out to at least 20 
kpc. Carignan (1985) also notes that NGC 247 has a comparatively 
low central surface brightness. The blue and red profiles of 
NGC 247 have very different behaviour at large radii; while 
the red profile defines a single exponential that extends over much of the galaxy, 
the blue profile flattens at large radii, suggesting a gradient in 
luminosity-weighted age in the outer disk. Davidge (2006) resolved bright 
blue stars at large radii in NGC 247, with MSTO ages between $\sim 
16$ Myr and $\sim 40$ Myr, confirming that a young component is present at large radii.

	The azimuthally-smoothed light profile constructed 
from the W1 image is shown in Figure 5, where a distance of 3.8 Mpc has been assumed. 
The W1 profile can not be characterised by a single 
exponential law, and there is no evidence of truncation. The W1 profile 
is similar in shape to the Carignan (1985) blue profile, in the sense that 
both profiles flatten at large radii. The blue and W1 light profiles 
are suggestive of a Type III (i.e. anti-truncated -- Pohlen et al. 2005) 
morphology, although we caution that the red profile from Carignan (1985) 
follows a single exponential profile. Type III profiles 
have been attributed to external drivers, such 
as interactions (e.g. Younger et al. 2007; Roediger et al. 2012). However, 
secular factors, such as the initial angular momentum 
of a system can also produce a Type III profile (Herpich et al. 2015).

\begin{figure}
\figurenum{5}
\epsscale{1.0}
\plotone{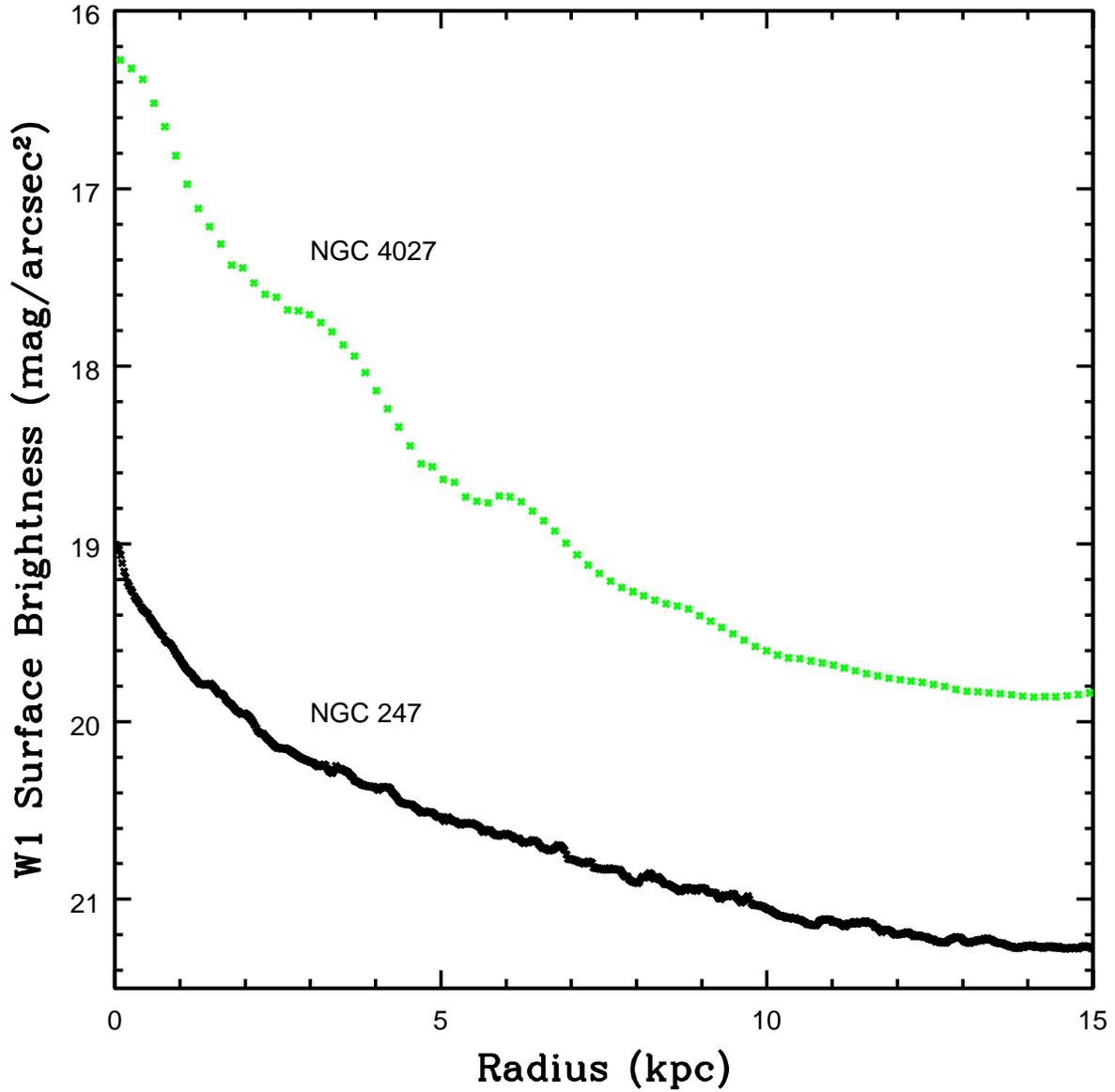}
\caption{Light profiles of NGC 247 and NGC 4027 that were obtained by 
azimuthally smoothing deprojected W1 images of both 
galaxies. The spatial scale along the x axis assumes 
distances of 3.8 Mpc for NGC 247 and 25.6 Mpc for NGC 4027. Neither profile 
can be characterized by a single exponential, and there is no evidence of truncation.}
\end{figure}

\section{THE ANGULAR DISTRIBUTION OF IR SOURCES}

	The brightnesses of luminous red objects in the NGC 247 disk 
can be measured from the SPITZER images discussed in Section 2, although 
the poor intrinsic resolution of the observations ($\sim 800$ parsecs$^2$ per 
resolution element at the distance of NGC 247) means that what 
appear to be individual sources are in reality blends. 
The certainty of blending notwithstanding, the light in individual resolution elements 
that do not contain large star clusters might be dominated by a single intrinsically 
bright star, and this expectation is consistent with the peak brightnesses 
in the NGC 247 CMD (see below). In any event, the identification of 
unblended sources is not critical for the comparative analysis that is conducted here. 

	The brightnesses of objects in the SPITZER images were measured with 
the point spread function (PSF)-fitting program ALLSTAR (Stetson 1994), 
with the PSFs, source catalogues, and initial photometric measurements 
that are used by ALLSTAR generated with routines in DAOPHOT (Stetson 1987).
The resulting $([3.6], [3.6]-[4.5])$ CMDs of stars in four radial 
intervals are compared in the top portion of Figure 6. 
The northern void is located in the 480 -- 660 arcsec interval.

\begin{figure}
\figurenum{6}
\epsscale{0.8}
\plotone{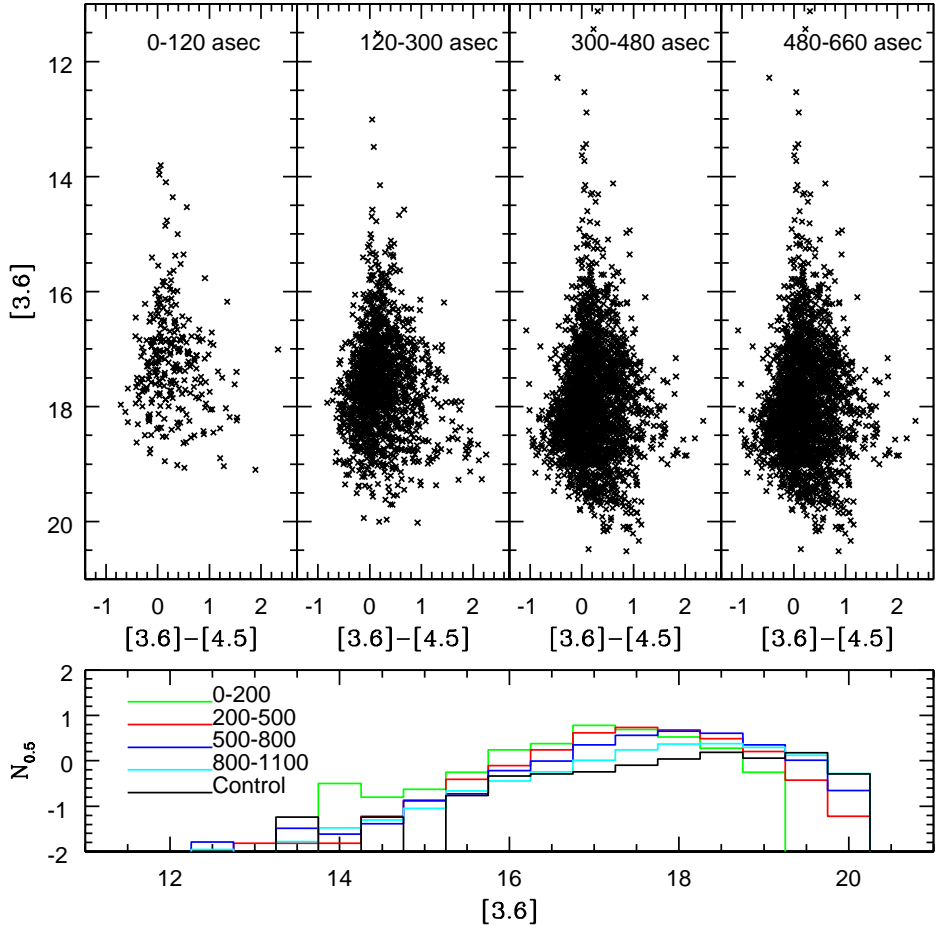}
\caption{$([3.6], [3.6]-[4.5])$ CMDs of stars in four radial intervals in 
NGC 247 are shown in the top row. The angular distances shown in each panel are 
measured from the galaxy nucleus in the deprojected images. Stars in 
the northern void are in the 480 -- 660 arcsec interval. The LFs 
of objects in these intervals, as well as that of a control field selected to monitor 
objects that are not members of NGC 247, are compared in the bottom panel, where 
N$_{0.5}$ is the number of objects per square deprojected arc minute per 0.5 
magnitude interval in [3.6]. The source counts for the innermost region are 
consistently higher than in the other three regions at the bright end, and these 
are likely unresolved star clusters and density flucuations 
in the central regions of the galaxy. The density of 
sources in all four fields exceed those in the control field near [3.6] = 17, 
suggesting that many of the objects with [3.6] $> 17$ belong to NGC 247.}
\end{figure}

	With the exception of a smattering of sources that 
may be dust-enshrouded stars or unresolved background 
galaxies, the majority of points in the CMDs define a vertical sequence. 
This vertical trajectory is due to the low temperature sensitivity 
of objects at MIR wavelengths that have effective temperatures that are 
characteristic of stellar photospheres (i.e. log(T$_{eff}) \gtrapprox 3$). The 
faint limit in the CMDs shown in Figure 6 moves to progressively fainter magnitudes 
as radius grows, reflecting the effect of stellar density on the faint limit.

	Foreground stars and background galaxies 
occur in significant numbers at the magnitudes examined in Figure 6. 
To estimate the angular densities of these contaminants, 
number counts were made in one corner of the SPITZER images where there is 
little or no contribution from bright sources belonging to NGC 247. The resulting 
number counts are compared with those in the four radial NGC 247 intervals 
in the lower panel of Figure 6. 

	The angular densities of objects in all four intervals exceed 
that in the control field when [3.6] $\sim 17$, indicating that [3.6] $= 17$ is the 
approximate peak of the CMD sequence throughout much of NGC 247, again with the caveat 
that many NGC 247 sources are expected to be blends. Blending should be 
most noticeable in the innermost annulus, and there is an excess number of sources with 
respect to foreground/background objects near [3.6] $\sim 16$ in that annulus.
These objects are intrinsically brighter than individual stars (see below), and 
are probably compact star clusters and/or stochastic density fluctuations. 
There is not an obvious transition signature in the CMDs in 
Figure 6 near [3.6] = $16 - 17$ due to the modest temperature sensitivity of 
the [3.6]--[4.5] color, although there may be an 
onset in the number of objects to the right of the vertical 
sequence at [3.6] magnitudes between 16 and 17.

	Blum et al. (2006) examine the IR properties of stars in the LMC, 
and their results provide benchmarks for interpreting the NGC 247 CMDs. 
Assuming a distance modulus of 18.5 for the LMC then 
red supergiants (RSGs) in the LMC have a peak absolute magnitude 
of M$_{[3.6]} = -10.5$, while AGB stars have M$_{[3.6]} = -9.5$. 
In NGC 247 these correspond to [3.6] = 17.5 for RSGs and [3.6] = 18.5 for AGB stars. 

	The spatial distribution of stars in 
the deprojected disk is examined in Figure 7. 
The comparisons made in the lower panel of Figure 6 
indicate that the majority of sources in the left hand panel are 
likely foreground/background objects, and this is consistent with the 
uniform distribution of sources in that panel that are more than a few 
tens of arc seconds from the nucleus. The modest number of sources 
in the left hand panel that are clustered near the nucleus of NGC 247 are 
likely unresolved star clusters and/or statistical flucuations. The more concentrated 
nature of sources in the right hand panel are consistent with many of the objects 
in the [3.6] = 17 -- 18 interval belonging to NGC 247.

\begin{figure}
\figurenum{7}
\epsscale{1.0}
\plotone{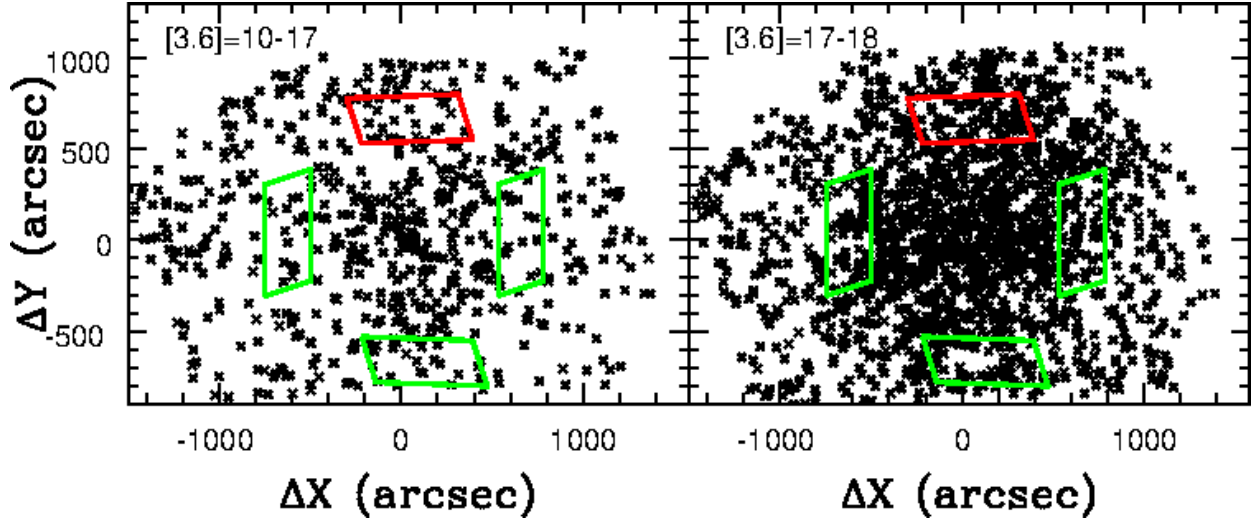}
\caption{Distributions of stars in two [3.6] magnitude intervals. The co-ordinates 
are in the deprojected disk, with (x,y) = (0,0) corresponding to
the galaxy nucleus. The red box marks the location 
of the void, while the green boxes mark areas where 
the comparison CMDs and LFs that are shown in Figure 8 were extracted. 
The more-or-less uniform distribution of objects 
in the left hand panel suggests that many of the objects with [3.6] $> 17$ 
are in the foreground or background, in agreement with the LF comparisons in 
Figure 6. There is a centrally concentrated 
body of objects in the right hand panel, indicating that there is a number 
of objects with [3.6] between 17 and 18 that belong to NGC 247, although 
foreground/background contamination is still significant.
The distribution of objects in the right hand panel suggests that the 
density of stars in the void is not markedly different from that at other radii, 
and this is consistent with the LF comparisons made in the bottom panel of Figure 8.}
\end{figure}

	The location of the northern void is indicated in both 
panels of Figure 7, as are the locations of three reference fields at 
the same galactocentric radius as the void. If the 
stellar content of the NGC 247 disk is well mixed 
then the incidence of blending should be the same in areas that have the same 
stellar density. Comparisons of photometric properties between areas that have 
similar surface brightnesses thus provide a means of assessing whether or not 
there are differences in the properties of the brightest stars even in the 
presence of blending. The distribution of stars in the right hand panel of 
Figure 7 indicates that the number of IR bright stars 
in the northern void is not remarkable when compared with 
other parts of the disk at similar galactocentric radii. This is consistent with 
the deprojected W1 image that was examined in the previous section, 
where the surface brightness of the void was found to be consistent with 
its location in NGC 247. 

	The $([3.6], [3.6]-[4.5])$ CMDs of sources in the northern void and the 
disk reference areas indicated in Figure 7 are compared in the top panel of Figure 8. 
There is a modest number of objects with [3.6]--[4.5] $\gtrapprox 1$ in all four CMDs 
with [3.6] $> 18$, and these are resolution elements where the light might be dominated 
by C stars. The number densities of these very red objects are similar in all 
four fields. 

\begin{figure}
\figurenum{8}
\epsscale{1.0}
\plotone{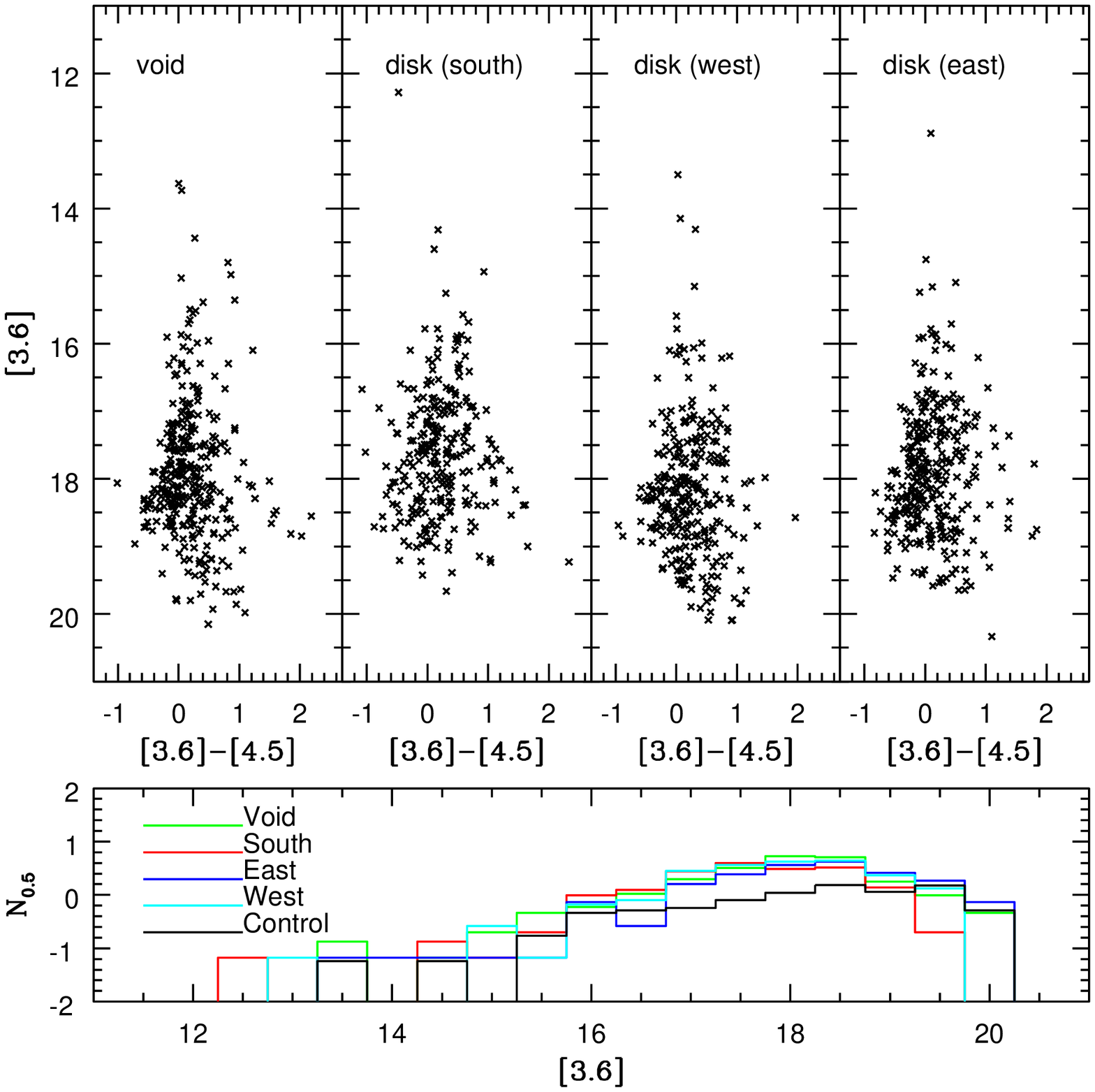}
\caption{$([3.6], [3.6]-[4.5])$ CMDs of stars in 
the northern void (left hand panel), and the 
reference areas indicated in Figure 7. The LFs of all 
four regions are compared with the LF of the foreground/background control 
field in the lower part of the figure, where N$_{0.5}$ is the number 
of stars per 0.5 magnitude interval in the [3.6] filter per square arc minute. 
The LFs of the four fields are similar when [3.6] $> 17$, 
and this is also where the star counts are consistently higher than 
in the control field. The number density of luminous red sources in 
the void agrees well with that in the disk reference fields.}
\end{figure}

	The LFs of stars in the void and the disk reference areas are 
compared in the lower panel of Figure 8, along with number counts from the control 
field. Comparisons with the control field are consistent with the main body of 
sources in the NGC 247 SPITZER images having [3.6] $> 17$, in agreement with that 
found in Figure 6 using larger areal coverage. In addition, while 
Wagner-Kaiser et al. (2014) find that there has not been recent star formation 
in the void during the past Gyr, the similarities between the LFs in 
Figure 8 suggests that the number density of stars in the northern void that formed 
during later epochs agrees with that in other parts of the disk at the same radius. 

\section{COMPARISONS WITH THE LOP-SIDED SPIRAL GALAXY NGC 4027}

	That NGC 247 is a lop-sided spiral galaxy 
has important implications for understanding the nature of the northern 
void. A lop-sided morphology opens the prospect that the northern void is the gap 
between the main body of the disk and the densely populated (when 
compared with other parts of the galaxy at the same 
galactocentric radius) northern spiral arm when seen in projection. 
Comparisons with the photometric properties of well-established 
late-type lop-sided spirals are then of obvious interest.

	NGC 4027 (Arp 22) is a lop-sided late-type barred spiral galaxy that 
is viewed almost face-on. NGC 4027 has the second-highest $m=1$ A$_1$ Fourier 
coefficient in the Zaritsky \& Rix (1997) $K-$band sample of lop-sided spiral galaxies, 
placing it in their 'significantly lop-sided' category. NGC 4027 is in the same 
group of galaxies as the interacting pair NGC 4038 and NGC 4039 (the 'Antennae'), and 
Phookun et al. (1992) detect a ring of gas around NGC 4027 that they attribute to an 
interaction with a companion. 

	NGC 4027 is a good reference object for the current study as it is 
viewed almost face-on, has an unmistakeable lop-sided character, is relatively 
nearby by cosmic standards, and has a body of high quality observations. We emphasize 
that the comparisons made here do not assume that NGC 247 and NGC 4027 have similar 
global properties (e.g. total mass, mass-to-light ratio, environment etc) and/or 
SFHs. Rather, we simply compare their appearances at wavelengths that 
sample the dominant contributors to stellar mass. 

	W1 and [3.6] images of NGC 4027 were downloaded from the IRSA website. 
The [3.6] images are the AOR 49579264 observations from 
SPITZER program 10046 (PI: D. B. Sanders). While sampling a similar 
range of wavelengths, the W1 and [3.6] images have different image qualities 
and photometric depths. Systematic effects that arise from the differences between 
these quantities simulate the impact of distance on the observations, which in turn 
affects intrinsic resolution within a galaxy and the brightnesses of sources 
that can be detected within a galaxy. NGC 4027 images in the FUV, NUV, and 
W3 filters are not considered since the present comparison 
is restricted to the global morphological properties of NGC 247 and NGC 4027 
as defined by the stars that contribute most to their stellar masses. 

	As with the NGC 247 images, the W1 
and [3.6] images of NGC 4027 were rotated to align the major 
axis with vertical. A smooth disk component was then constructed by azimuthally 
averaging the light profile about the nucleus, and the result was subtracted from the 
downloaded images. The initial rotated and disk-subtracted images of NGC 4027 
are shown in the left hand and central columns of Figure 9. 

\begin{figure}
\figurenum{9}
\epsscale{1.0}
\plotone{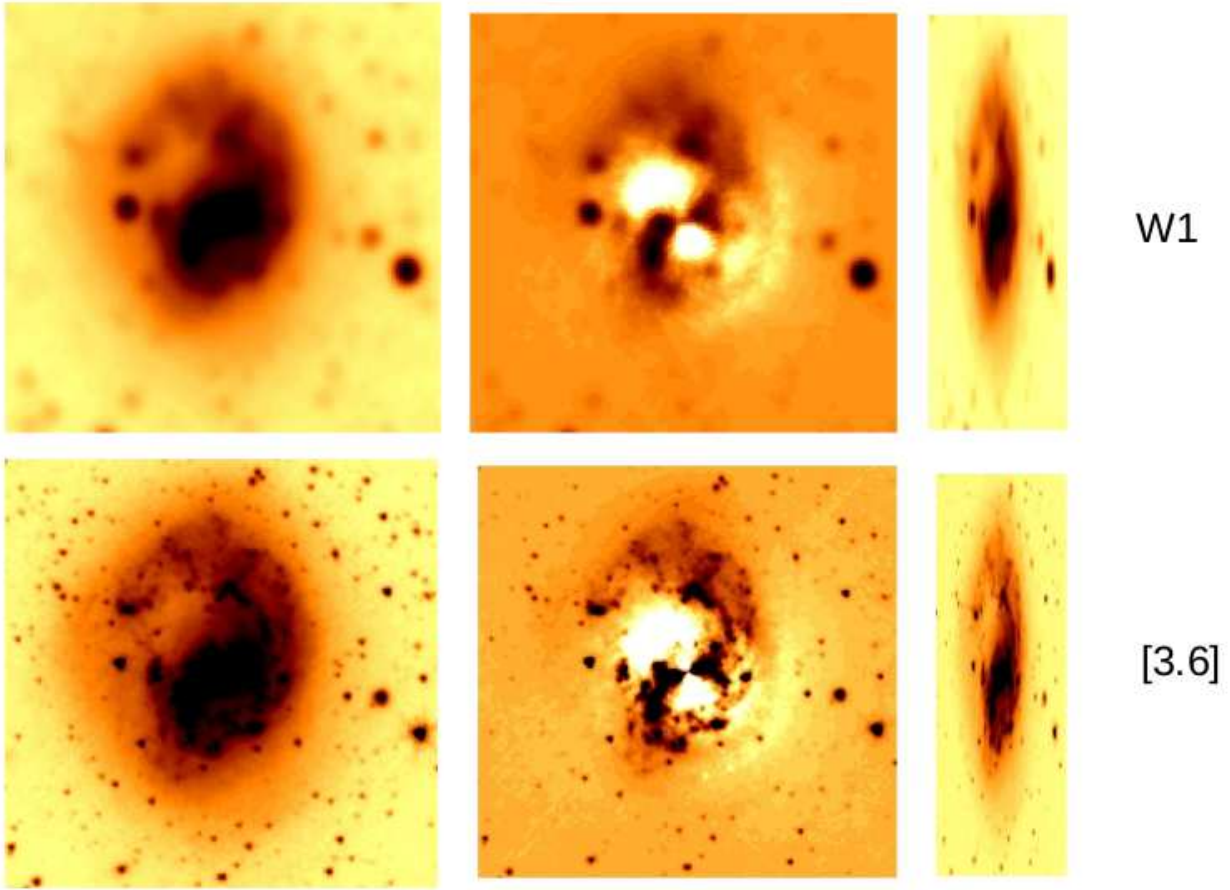}
\caption{W1 and [3.6] images of NGC 4027, 
rotated so that the major axis of the galaxy is aligned with vertical. 
The initial rotated images are shown in the right hand column, while the 
middle column shows the results of subtracting an azimuthally-smoothed disk 
component. Differences due to image quality and exposure time are such that 
the W1 image simulates NGC 4027 as it would appear if viewed by SPITZER in [3.6] 
at a distance that is $4 - 5\times$ greater than its actual value. 
The right hand panel shows NGC 4027 as it would appear if viewed 
with the same line-of-sight inclination as NGC 247. The apparent 'void' in the NGC 
4027 disk when the galaxy is tilted to the line of sight is the 
region between the main body of the disk and the perturbed spiral arm.}
\end{figure}

	The disk-subtracted W1 and [3.6] images of NGC 4027 are broadly 
similar, suggesting that distance effects should 
not be an issue when making comparisons with NGC 247. Of greater 
importance for the present work is that there are 
clear similarities between the deprojected and disk-subtracted images of NGC 
4027 and their NGC 247 counterparts in the top row of Figure 2. As with NGC 247, 
there is a prominent spiral arm along the major axis of NGC 4027 
that stands out after disk subtraction. There is also residual structure 
near the center of the disk-subtracted image of NGC 4027 that is similar to that 
near the center of NGC 247. The area between the upper spiral arm and the main body 
of NGC 4027 -- which corresponds to the northern void in NGC 247 -- has negative 
residuals in the disk-subtracted image, as is seen in the corresponding NGC 247 image.

	The NGC 4027 images were also compressed to simulate 
its appearance if viewed with the same line of sight inclination as NGC 247, and 
the results are shown in the right hand column of Figure 9. There is good overall 
agreement between these NGC 4027 images and their NGC 247 counterpart in Figure 2. 
The space between the upper spiral arm and the main body of the 
NGC 4027 disk is similar in appearance to the void in NGC 247. 

	The NGC 4027 light profile that was obtained by azimuthally smoothing the W1 
image is shown in Figure 5. While there is a clear offset in surface 
brightness, there are clear similarities with the shape of the NGC 247 
light profile. The NGC 4027 light profile can not be characterized by a single 
exponential, and extends out to large radii without signs of truncation.

\section{DISCUSSION \& SUMMARY}

	Archival UV and IR images have been used to examine the structure 
of the late-type spiral galaxy NGC 247. This galaxy is of particular 
interest as (1) it is a companion of the starburst galaxy NGC 253, and (2) it has 
been suggested that there was an interaction with a dark halo within the past 
Gyr that scoured gas from a part of its disk. The light in the 
images examined in this paper originates from stars that span 
a range of stellar ages from the youngest (FUV, NUV, and W3 images) to 
the oldest (W1 images) stars. The images have been deprojected to examine the 
face-on appearance of the galaxy. Disk light that is uniformly distributed in the 
azimuthal direction has also been removed to enhance asymmetries. 

	The main conclusions are as follows:

\vspace{0.3cm}
\noindent{1)} The surface brightness of the northern spiral arm of NGC 247 
exceeds that in other parts of the disk at the same galactocentric radii. This 
applies over a wide range of wavelengths, suggesting that the northern spiral arm 
contains a comparatively high number density of stars that span a range of 
ages. This does not mean that the northern spiral arm is an old structure, 
but simply that it contains stars that formed over a range of ages. The age of 
this structure notwithstanding, it gives the disk of NGC 247 a highly 
lop-sided appearance when viewed face-on.

\vspace{0.3cm}
\noindent{2)} While the incidence of blending is high in the SPITZER 
images, it is still possible to conduct a differential 
comparison between areas that have similar stellar densities. 
The density of luminous red sources in the northern void of NGC 247 
measured from the SPITZER [3.6] and [4.5] images agrees with that in other parts 
of the galaxy at similar galactocentric radii. This agreement 
applies not only to the total number of detected sources, but also to 
LFs. This agreement is consistent with the void being the gap between the northern 
spiral arm and the main body of the disk in a lop-sided system (see below). 
Blending issues aside, a detailed assessment of the SFH based on 
these data is complicated by stellar variability among luminous evolved stars, which 
blurs age sensitivity at these wavelengths (Davidge 2014). 

\vspace{0.3cm}
\noindent{3)} Prominent bubbles are seen in the UV images. One is near the 
base of the northern spiral arm, while a second is in the southern disk. 
Light from these bubbles likely originates from young stars and nebular emission, 
and their presence hints at wide spread star formation within the past Gyr. Indeed, 
dynamical ages have been estimated for these bubbles by adopting expansion velocities 
measured for similar structures in the disk of Holmberg II by Puche et al. (1992), and 
the results coincide with a recent episode of elevated levels of star formation 
in the nucleus of NGC 247. We speculate that the luminous sources in the 
central few kpc of NGC 247 that are seen in the W3 image 
are luminous dust-enshrouded AGB stars that also formed at this time.

	While large-scale bubbles are usually attributed to energy injected into the 
ISM by star-forming activity, Mirabel \& Morras (1990) find that 
the accretion of HVCs by the Galaxy can inject significant amounts of 
energy into the ISM that can be comparable to what is needed to form supershells 
(e.g. Park et al. 2016). If this is the case then the area of highly active star 
formation in the NGC 247 may have been localized near the areas affected by HVCs 
within the past Gyr.

\vspace{0.3cm}
\noindent{4)} Images of NGC 247 that are processed to simulate a face-on orientation 
show similarities with the classic lop-sided spiral galaxy NGC 4027 (Arp 22). Not 
only do NGC 247 and NGC 4027 have similar asymmetric structure, but they also have 
azimuthally smoothed light profiles with similar shapes. The gap between the dominant 
spiral arm and the main body of the disk in both galaxies is 
deficient in light in W1 after light from a smooth disk component is removed.
Given the evidence that NGC 247 is a lop-sided spiral galaxy then 
the northern void is actually the gap between the high 
surface brightness northern spiral arm and the main body of the disk.

	NGC 4027 is an example of a 'significantly' lop-side spiral galaxy using 
the structural criterion discussed by Zaritsky 
\& Rix (1997), and the comparisons made here indicate that this 
characterization also holds for NGC 247. We emphasize that similarity 
in appearance does not imply that NGC 247 and NGC 4027 have had similar SFHs or 
have been subject to similar evolutionary processes. Lop-sided 
structures can result from a number of factors (see discussion 
below), and NGC 247 and NGC 4027 are in very different environments. 

	Rudnick et al. (2000) find comparatively high SFRs in lop-sided systems. 
While the apparent low present-day SFR in NGC 247 appears to run counter to this, 
the quenching timescale for star bursts in low mass galaxies is $0.5 - 1.3$ Gyr 
(e.g. McQuinn et al. 2010). It is then possible that elevated levels of star formation 
in NGC 247 within the past few hundred Myr may have been recently quenched. 

	Lop-sided spiral galaxies are not rare (e.g. Bournard et al. 2005; Reichard 
et al. 2008), suggesting that the formation of the asymmetries that define these 
galaxies is somehow tied to events that are common in disk galaxy evolution (Zaritsky 
et al. 2013). A number of mechanisms have been forwarded to explain the origins of 
lop-sided disks (e.g. Reichard et al. 2009, Zaritsky et al. 2013 
and references therein), and these can be categorized as external (e.g. tidal 
interactions, mergers, and the cosmological accretion 
of gas) and internal (e.g. secular evolution and asymmetries in the halo) in nature. 
Zaritsky et al. (2013) suggest that weak lop-sided structure is tied to internal 
origins, such as halo asymmetries. If this is correct then the large-scale 
nature of the asymmetric disk of NGC 247 makes it likely that its lop-sided nature 
is external in origin. Therefore, two external mechanisms (tidal interactions 
and gas accretion) are discussed here to assess their viability as 
possible causes of the lop-sided nature of NGC 247. 

	Zaritsky \& Rix (1997) discuss evidence of a 
tidal origin for asymmetric arms. Tidal arms likely 
arise from an interaction during the past Gyr (i.e. within a few disk rotation 
times), and tend to have elevated SFRs (e.g. 
Holincheck et al. 2016) when compared with the main body of the 
host galaxy. The latter is consistent with the FUV and IR observations of 
the northern spiral arm in NGC 247 that indicate it is an area of concentrated 
star-forming activity. The transfer of angular momentum 
due to an interaction could also form an extended disk 
(e.g. Hammer et al. 2007), with a Type III light profile 
(e.g. Younger et al. 2007; but see also Herpich et al. 2015), as might be 
the case for NGC 247 (Section 3). 

	If the northern arm in NGC 247 is tidal in origin then where is 
the perturbing galaxy? Arguably the most obvious candidate 
is the well-studied starburst galaxy NGC 253. 
NGC 247 and NGC 253 are close together physically 
(Karachentchev et al 2003), and have angular momentum vectors that are 
consistent with a common origin (Whiting 1999), and/or a history 
of angular momentum exchange. An interaction between NGC 247 and NGC 253 may have 
triggered the elevated levels of star formation that are seen at this day in the latter 
galaxy. In fact, the timing of the uptick in the nuclear and 
circumnuclear SFR of NGC 247 a few hundred Myr in the past found by Kacharov 
et al. (2018), and the ages estimated for bubbles in NGC 247 in Section 3 more-or-less 
coincide with the timing of the onset of the NGC 253 star burst (e.g. Davidge 2010). 
As the nearest galaxy of comparable size to NGC 247, NGC 253 may also have been 
a donor of chemically-enriched material for star formation, thereby 
accounting for the higher than expected metallicity 
of recently formed stars found by Kacharov et al. (2018). 

	There are problems with a recent tidal interaction between NGC 247 and 
NGC 253. Perhaps the most significant of these is that a debris field would be a likely 
outcome of such an event. However, a debris field between NGC 247 and NGC 253 has 
yet to be detected.

	There are alternatives to a close encounter with NGC 253 that could 
explain a tidal origin for the northern spiral arm in NGC 247. 
NGC 247 and NGC 253 are accompanied by an entourage of smaller 
companion galaxies, and most of these appear to be centered around NGC 247 
(Karachentsev et al. 2003). However, these galaxies are modest in size, and it is 
unlikely that they could induce a large scale tidal feature like the northern arm. 

	NGC 247 may also have been perturbed during a flyby 
encounter with another galaxy. The identification 
of the perturbing galaxy in such an encounter is problematic given that the 
interaction happened hundreds of Myr in the past. While the Sculptor group 
has considerable depth along the line of sight, most of its galaxies are within a 
Mpc of NGC 247 (Karachentsev et al. 2003). Given that NGC 247 and NGC 253 
are physically close then a flyby might even have triggered elevated SFRs in both 
galaxies. 

	Bournaud et al. (2005) point out that tidal 
interactions can not explain all of the properties of lop-sided 
spirals, and suggests that asymmetries may also be induced by the accretion 
of cosmological gas. Dupuy et al. (2019) discuss 
the accretion of HVCs onto barred spirals, and conclude 
that they can produce low amplitude lop-sided structure. 
This conflicts with the large-scale nature of asymmetries in the NGC 247 disk. 
Reichard et al. (2009) also note that the events that drive lop-sidedness tend to 
involve the inflow of low metallicity gas, and this is contrary to what has been 
inferred for recently-formed stars near the center of NGC 247 by Kacharov et al. (2018).

\acknowledgements{It is a pleasure to thank the anonymous reviewer for providing a 
prompt and helpful report.}

\appendix

\section{Deconvolution Experiments}

	The deprojection procedure decribed in Section 3 elongates the PSF in images, 
smearing light along one axis. To examine the effect of this smearing on diffuse 
sources in the disk of NGC 247, Wiener filters were applied 
to the deprojected NUV image to correct for this distortion. 
One filter (`aggresive Wiener') was designed to reproduce a 
Gaussian PSF with a full width at half maximum (FWHM) that matches the PSF in the 
original (i.e. with no deprojection) image. Such a filter greatly amplifies 
noise in the frequency spectrum along the elongated axis. A less ambitious 
filter (`moderate Wiener') was also generated to produce a Gaussian PSF with a 
FWHM that is midway between that of the original and elongated PSF.

	The deconvolved images are shown in Figure A1. The 
low-frequency response function of the filters produce streaks that become 
evident when the filter is convolved with very bright sources. The streaks are along 
the horizontal axis as that is the orientation of the distortions that the 
filters are designed to correct. The filter response artifacts extend over large 
angular scales as the low frequency component of the signal must be altered 
substantially to produce a symmetric Gaussian PSF from the elongated PSF.
Ringing is also seen in the image produced with the aggressive 
Wiener filter. All in all, the artifacts of the 
deconvolution procedure highlight the pitfalls of extracting information from 
portions of the frequency spectrum that have a low S/N ratio. 

	By design, the PSFs of stars in the deconvolved images are round. 
The noise introduced by the filters aside, the structures 
seen in the initial image are largely unchanged in the deconvolved images. The 
deprojection procedure thus has not greatly distorted structures in the 
NGC 247 disk.

\begin{figure}[!ht]
\figurenum{A1}
\epsscale{1.0}
\plotone{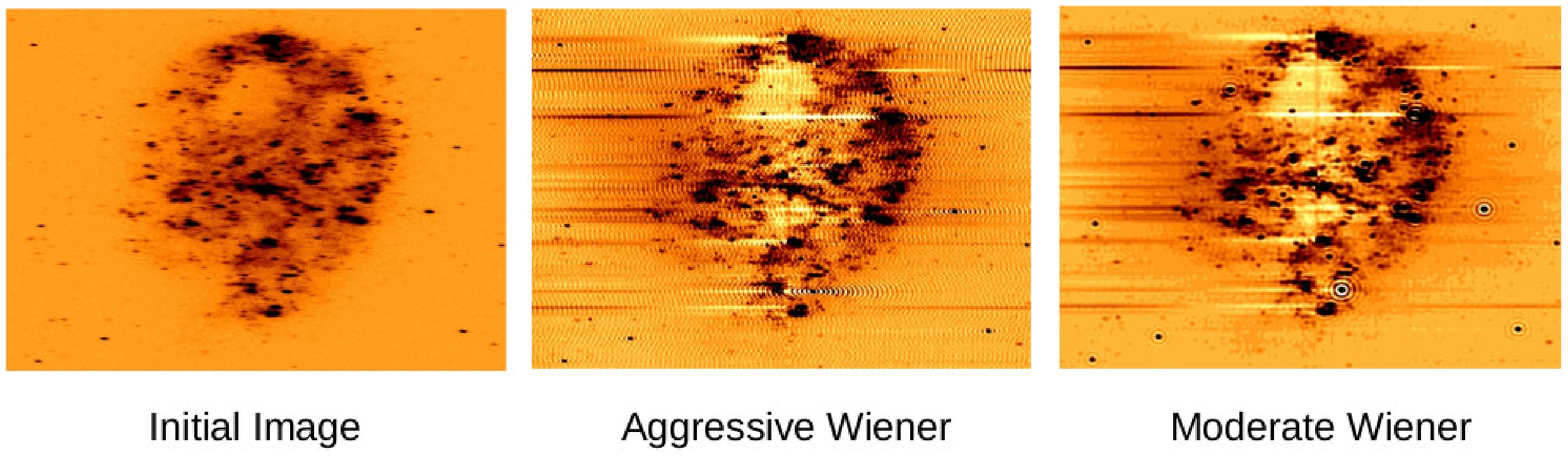}
\caption{Results of the application of the two Wiener filters described in the text 
to the initial NUV image. The filters collapse the elongated PSFs 
that are the result of the deprojection process. 
The horizontal streaks are the low frequency component 
of the filter response functions that are highlighted when the filter is 
convolved with bright sources, and these could be suppressed with additional 
filtering. That the shapes of structures in 
the deconvolved images are similar to those in the unfiltered image 
demonstrates that the deprojection process has not greatly distorted features in 
NGC 247.}
\end{figure}

\end{document}